\documentclass[12pt]{article}

\usepackage{amsmath}
\usepackage{epsf}
\begin{document}
\sloppypar

 {\it Accepted for publication in Astronomy
Letters, v.  26, N 12, 2000.}
\vspace{2cm}
\bigskip
 
\large
\centerline{\bf Energy Release on the Surface of a Rapidly Rotating Neutron}
\centerline{\bf Star during Disk Accretion: A Thermodynamic Approach.}
\vspace{15mm}
 \normalsize
\centerline{ Nail Sibgatullin$^{1,2}$ and Rashid A. Sunyaev$^{2,3}$}
\vspace{2mm}
\noindent
$^1${\it Moscow State University, Vorob'evy gory, Moscow, 119899 Russia }\\
$^2$ {\it Max-Planck-Institut f\"ur Astrophysik,
Karl-Schwarzschild-Str. 1, 85740 Garching bei \\
 \noindent Munchen, Germany}\\
$^3$ {\it Space Research Institute, Russian Academy of Sciences, ul. 
Profsoyuznaya 84/32, Moscow, 117810 Russia } \\
$^*$ {e-mail:sibgat@mech.math.msu.su}
 
\vspace{7mm}
\def\f{\frac}
\def\p{\partial}
{\small ...Abstract. }
The total energy $E$ of a star as a function of its angular 
momentum $J$ and mass $M$ in the Newtonian theory: $E = E(J, M)$ [in 
general relativity, the gravitational mass $M$ of a star as a function of 
its angular momentum $J$ and rest mass $m$, $M = M(J, m)$], is used to 
determine the remaining parameters (angular velocity, equatorial radius, chemical 
potential, etc.) in the case of rigid rotation. Expressions are derived 
for the energy release during accretion onto a cool (with constant entropy), rapidly rotating neutron star (NS) in the Newtonian theory and 
in general relativity. A separate analysis is performed for the cases 
where the NS equatorial radius is larger and smaller than the radius of 
the marginally stable orbit in the disk plane. An approximate formula is 
proposed for the NS equatorial radius for an arbitrary equation of 
state, which matches the exact one at $J = 0.$


\pagebreak

\section*{Introduction}

The change in mass and angular momentum of a cool neutron star (NS) 
during accretion leads to a transformation of its equilibrium state. 
Below, we consider NSs with weak magnetic fields that do not affect the 
accretion dynamics; i.e., our results refer to the accretion pattern in 
low-mass X-ray binaries and, in particular, in X-ray bursters. Of 
crucial importance is the question of what part of the energy released 
during accretion dissipates near and outside the stellar surface (in the 
accretion disk, in the boundary layer, in the spread layer, in the 
settling zone as matter is compressed under the weight of the newly 
supplied material, when uniform rotation is established by viscous 
forces throughout the entire extended stellar atmosphere, in the zone 
with a surface density of the order of $10^9 g/ cm^2$ subject to nuclear 
burning during X-ray flares) and what part dissipates inside the star in 
its interior. Observationally, these two zones of energy release differ 
radically. The energy released near the surface leaves the surface 
layers (is emitted) in a very short time: from fractions of a 
millisecond to several tens of seconds, although the complete energy 
dissipation can last for several hours. By contrast, the energy release 
in the stellar interior produces radiation with characteristic times 
exceeding hundreds of thousands and, possibly, millions of years; this 
is the time it takes for the interiors of neutron stars with weak 
magnetic fields to cool down (Levenfish et al. 1999). Such radiation 
can be detected only in highly variable transients when accretion on 
them virtually ceases.

Here, we disregard energy release in the stellar interior by assuming 
that the star is cool and that its entropy does not change during 
accretion and transformation of the NS internal structure. This strong 
assumption allowed us to obtain a number of general results, which we 
used previously (Sibgatullin and Sunyaev 2000). Note that this 
assumption does not hold in the so-called thermal neutron stars, where 
the spinup via accretion leads to a difference between the angular 
velocities of the crust and the central liquid superfluid core inside 
the star. As a result, energy is released inside the star through 
viscous friction [see Alpar (1999) for a discussion].

A thermodynamic relation between the change in NS total energy 
(gravitational mass in general relativity) and the change in its angular 
momentum and mass (rest mass in general relativity) is considered for 
cool, rigidly rotating NSs with a given equation of state (EOS). This 
relation and the energy conservation law are used to derive a formula 
for the energy release on the stellar surface (in the Newtonian 
approximation and in general relativity) during disk accretion onto a NS 
rotating with an arbitrary angular velocity. The case where the NS 
equatorial radius is smaller than the radius of the marginally stable 
orbit is also considered in terms of general relativity.

For the equilibrium figure of a rotating, incompressible fluid in its 
own gravitational field (Maclaurin spheroid), we derive explicit 
formulas for the dependence of disk and surface energy release on 
spheroid eccentricity. The fraction of disk energy release in the total 
energy release is shown to be expressed by a simple linear dependence on 
the ratio of the Maclaurin-spheroid rotation frequency to the Keplerian 
equatorial particle velocity.

A universal approximate formula for the NS equatorial radius is derived 
for an arbitrary EOS.

Previously (Sibgatullin and Sunyaev 2000), we considered astrophysical 
implications of our results and methods for deriving simple 
approximation formulas. We also analyzed the universal geometric 
properties of space-time outside rotating bodies.

\section{Newtonian treatment}

In the Newtonian approximation, the total energy of a NS consists of the 
gravitational, kinetic, and internal energies:
\begin{equation}
{\bf E}=\int_{V}(-\frac12 \Phi(\overline{r})
+\frac12\Omega^2(x^2+y^2)+u(\rho(\overline{r})))\rho(\overline{r}) \,dV
\end{equation}
where the gravitational potential $\Phi$   is
\begin{equation}
\Phi(\overline{r}) = G \int_{V}\frac{\rho(\overline{r}\prime)}{\mid \overline{r}-\overline{r}\prime\mid} \,dV^{\prime}
\end{equation}
We take a model of an ideal gas with constant (zero) entropy for the 
internal energy of superdense matter. The Gibbs identity then yields
\begin{equation} d u= -p
d(1/\rho) + Td s = -p
d(1/\rho).
 \end{equation}
We have the following obvious expressions for the NS mass and angular 
momentum:
\begin{equation}
M=\int_{V}\rho(\overline{r})\,dV,\qquad J =
\int_{V}\rho(\overline{r})(x^2+y^2)\Omega\,dV
\end{equation}
The following integral holds in a steady equilibrium state:
\begin{equation}
u(\rho) +\frac{p}{\rho}-\Omega^2(x^2+y^2)/2 -\Phi=\mu= \mbox{const},
\end{equation}
since  $\mu$ has the same value at any point of the star \footnote{$^1$ 
The constant $\mu$ should not be confused with the Bernoulli integral 
$i_0,$ which is constant along a streamline: $\mu = i_0 - [r, \Omega]^2$ !}, 
because the equilibrium is isentropic, and because the rotation is rigid 
[the theorem of Crocco (1937); see also Oswatitsch (1976), Chernyi 
(1988)]. So, the dynamical equilibrium conditions can also be written as 
$\nabla\mu = 0$ [see the case of an incompressible fluid in Lamb (1947)].

Let us consider two close equilibrium states with global parameters  ${\bf E}, M, J, \Omega$ and ${\bf E} +
\delta{\bf E}, M + \delta M, J +\delta J, \Omega +\delta \Omega$.

{\it Theorem 1.$\quad$ The variations of NS mass $\delta M$ , angular momentum 
$\delta J$, and total energy $\delta {\bf E}$ for two close equilibrium states 
are related by the thermodynamic relation
\begin{equation}
\delta{\bf E} = \Omega\delta J + \mu\delta M
\end{equation}
}
Indeed, denote the local displacement of point $\overline{r}$ on the stellar surface along the normal when passing from one equilibrium state to the other by 
$W(\overline{r})$. For an arbitrary integral $A\equiv\int_{V} a(\overline{r})\,dV$, we then have
\begin{equation}
\delta A=\int_{V}\delta a(\overline{r})\,dV + \int_{\partial
V}a(\overline{r})W(\overline{r})\,dS
\end{equation}
In view of Eq. (2) for the gravitational potential, the variation of 
gravitational energy $\delta {\bf E}_{gr}$ is
$$
\delta {\bf E}_{gr} = -\f12 \int_{V}\delta\rho\Phi\,dV-\f12\int_{V}\rho\delta\Phi\,dV -\f12\int_{\partial
V}\rho W\Phi\,dS. 
$$
Reversing the order of integration in the second term on the right-hand 
side, we reduce this expression to
\begin{equation}
\delta {\bf E}_{gr} = - \int_{V}\delta\rho\Phi\,dV-\int_{\partial
V}\rho W\Phi\,dS
\end{equation}
We calculate the variation of internal energy $\delta {\bf E}_{in}$ from formula (7) by 
using the Gibbs identity (3):
\begin{equation}
\delta {\bf E}_{in} =\delta\int_{V}u(\rho)\rho\,dV=\int_{V}(u(\rho)+\frac{p(\rho)}{\rho})\delta\rho\,dV+
 \int_{\partial V}\rho W (u(\rho)+\frac{p}{\rho}) \,dS
\end{equation}
In order to simplify the subsequent calculations, we added the surface 
term $ \int_{\partial V} W p \,dS$ to the right-hand part of (9); this term is zero, because 
the pressure vanishes at the stellar boundary. Clearly, the variation of 
kinetic energy can be expressed as
$$
\delta{\bf E}_{Ó} =\f12 \int_{V}(\delta\rho\Omega +
2 \delta \Omega\rho)(x^2+y^2)\Omega\,dV +\f12\int_{\partial
V}\Omega^2\rho(\overline{r})(x^2+y^2) W\,dS.
$$
As a result, the variation of total energy $ \delta {\bf E}=\delta{\bf E}_{gr}+\delta{\bf E}_{in}+\delta{\bf E}_c$  is
\begin{eqnarray*}
&\delta{\bf E} =\int_{V}\left\{\delta\rho\left(-\Phi + \frac
12\Omega^2(x^2+y^2)+u(\rho) + p/\rho \right)+ \delta \Omega\Omega
(x^2+y^2)\right\}\,dV+\\
& \int_{\partial
V}\rho W\left(-\Phi + u(\rho) + p/\rho + \frac 12\Omega^2(x^2+y^2)\right)\,dS.
\end{eqnarray*}
$\delta M$, $\delta J$ are given, respectively, by
$$
\delta M =\int_{V}\delta \rho(\overline{r})\,dV + \int_{\partial
V}\rho(\overline{r})W(\overline{r})\,dS;$$
\begin{equation}
\delta J
 =\int_{V}(\Omega\delta \rho (\overline{r})+\delta\Omega \rho
(\overline{r}))(x^2+y^2)\,dV + \int_{\partial V}\Omega\rho(\overline{r})(x^2+y^2) W\,dS.
\end{equation}
Regrouping the terms, we represent the expression for $\delta{\bf E}$ as
\begin{eqnarray}
&\delta {\bf E} =\int_{V}\left\{\delta\rho\left(-\Phi - \frac
12\Omega^2(x^2+y^2)+u(\rho)+\frac{p}{\rho}\right)\right\}\,dV+\nonumber
\\
&\int_{V}\left\{\delta\rho\Omega+
 \delta \Omega\rho\right\}(x^2+y^2)\Omega\,dV +\int_{\partial
V}\Omega\rho(\overline{r})(x^2+y^2) W\,dS+\nonumber
\\
& \int_{\partial
V}\rho W\left(-\Phi +u(\rho)+\frac{p}{\rho}- \frac 12\Omega^2(x^2+y^2)\right)\,dS.
\end{eqnarray}
The coefficients in front of $\delta \rho$ in the first and last 
integrals of Eq. (11) match the constant $\mu$ given by (5). Let us now 
make use of Crocco's theorem and factor this constant outside the 
integral signs. The angular velocity $\Omega$  can also be factored 
outside the integral signs in the second and third integrals, because 
the rotation is rigid. Having done these operations and using formula 
(7) for $\delta M$ and $\delta J,$ we obtain
\begin{equation}
\delta {\bf E} = \Omega\delta J + \mu\delta M,
\end{equation}
hence,
\begin{equation}
\Omega=\f{\p
{\bf E}}{\p J}|_{M},\quad \mu = \f{\p
{\bf E}}{\p M}|_{J}.\tag{$12a$}
\end{equation}
We see that the constant $\mu$ has the physical meaning of chemical 
potential here (Landau and Lifshitz 1976).

Theorem 1 is conceptually associated with the following remarkable 
variational principle of general relativity by Hartle and Sharp (1967) 
(see also Bardeen 1970): the true mass and angular-momentum 
distributions differ from their virtual (possible) distributions with a 
fixed rest mass and a fixed total angular momentum in that they give a 
conditional extremum to the gravitational mass (energy in the Newtonian 
approximation). In this case, the angular velocity $\Omega$ and the 
constant $\mu$ act as the Lagrangian factors.

{\it Corollary 1. $\quad$ When a star loses its angular momentum and energy by the 
radiation of electromagnetic or gravitational waves, the rates of change 
in its total energy and angular momentum are related by $\dot{{\bf E}} = \Omega \dot{J}.$}

This equality follows from (12), because the change in NS baryonic mass 
is zero during wave emission. Corollary 1 was proved by Ostriker and 
Gunn (1969) when considering the fluxes of angular momentum and energy 
of electromagnetic or gravitational radiation in the wave zone. It has 
important applications for radio pulsars: the quasi-equilibrium 
evolution of the NS structure when it loses its angular momentum does 
not lead to any heating of the stellar matter.

Denote the fluxes of angular momentum and energy on the NS by $\dot{M}l $  and 
$\dot{M} e $, respectively, where $l$ and $e$ have the meaning of the specific 
angular momentum and specific energy brought by the accreting matter.

{\it Theorem 2. \quad When the angular momentum $\dot{M}l $ and energy $\dot{M} e $, are 
transferred to a neutron star by accreting particles in unit time, the 
following energy is released in the star with constant entropy in unit 
time:
\begin{equation}
L_s=\dot{M}( e  -\Omega\, l -\mu),
\end{equation}
where $\mu$ is the chemical potential of the cool star.}

Indeed, according to theorem 1, we have
$$
d {\bf E} = \Omega dJ +\mu dM.
$$
On the other hand, it follows from the energy conservation law that
$$
d{\bf E } = dM e  - L_sdt,
$$
where $L_s$ is the rate of energy release (stellar luminosity) during 
accretion.

From the law of conservation of angular momentum, we have
\begin{equation}
dJ =\dot{M} l dt.
\end{equation}
By equating the expressions for $d E$ from theorem 1 and from the energy 
conservation law and using (14) for $d J$, we reach the conclusion of 
theorem 2.

 {\it Note.} For the constant $\mu$, we may choose its value at the stellar 
equator: $\mu =
 -\frac12\Omega^2 R^2-\Phi_e$, where $\Phi_e  $ is the gravitational potential at 
the stellar equator, and $R$ is the equatorial radius. Here, we make use 
of the fact that the enthalpy vanishes on the stellar surface. This 
choice of  $\mu$ allows us to determine the equatorial radius, the most 
important NS parameter [see formulas (32) and (33)]. 

{\it Corollary 1. \quad When a thin accretion disk is adjacent to the stellar 
equator and when the star transforms into an equilibrium state with a 
new mass and angular momentum in time $dt,$ the energy $\frac12 dM (\Omega_K
-\Omega)^2 R^2$
is released on the stellar surface, where  $dM$ is the amount of baryonic 
mass accreted onto the NS in time $dt$; $\Omega_K$  and $\Omega$ are the Keplerian 
velocity at the NS equator and its angular velocity, respectively.}

Indeed, in this case, $ e = v^2/2-\Phi_e;\quad v= R \Omega_K, l =
R^2\Omega_K$, and 
expression (13) for $L_s$ turns into a full square. In particular, it 
follows from corollary 1 that the energy release for counterrotation at 
$\Omega =- 0.5\Omega_K$ is a factor of 9 larger than the energy release for 
corotation at $\Omega = 0.5\Omega_K$!

The local justification here is as follows. Falling on the stellar 
equator, a particle of mass $m$ increases the NS moment of inertia. An 
additional work is done, which is equal to the difference between the 
particle angular momenta in the Keplerian orbit, $m R^2\Omega_K,$ and on the 
stellar surface, $m R^2\Omega,$ multiplied by the NS angular velocity $\Omega.$ 
This work is equal, with an opposite sign, to the additional 
(rotational) energy that must be added to the difference between the 
particle kinetic energies in the Keplerian orbit, $m R^2\Omega_K^2/2$, and on the 
stellar surface,$m R^2\Omega^2/2$. Therefore, when a particle approaches the NS
from a thin accretion disk and decelerates in a narrow boundary layer at 
the equator from the Keplerian angular velocity $\Omega_K$ to the NS rotation 
velocity $\Omega$, it releases the energy $1/2 m(\Omega_K-\Omega)^2 R^2,$ giving up its angular 
momentum and part of its energy to the star. The formula for the energy 
release in this form was justified by Kluzniak (1987). The various local 
derivations of this formula were discussed by Kley (1991), Popham and 
Narayan (1995), and Sibgatullin and Sunyaev (1998).

It should be noted that, having accreted at the equator, the matter 
cannot remain there for long. It must spread in some way over the 
surface while changing its angular momentum and energy. Therein lies the 
inconsistency of the local approach, which disregards the subsequent 
redistribution of accreting matter over the star. The close match 
between the energy release given by (13) and the energy released when 
particles decelerate from the velocity $R\Omega_K$ to the velocity $R\Omega$ on the 
stellar surface is fairly unexpected.

Here, we make an attempt to solve the problem by taking into account the 
restructuring of a cool star during accretion. This global approach 
leads to formula (13). Thus, (13) holds not only for a narrow boundary 
layer near the equator (Popham and Sunyaev 2000). Our derivation of 
corollary 1 from theorem 1 suggests that the same energy is released by 
a particle in the spread layer (Inogamov and Sunyaev 1999), although the 
situation in the spread layer differs radically from the problem with a 
thin equatorial boundary layer. In the surface layer, the matter heats 
up, the thermal energy radiates away in bright latitudinal rings, and 
other complex physical processes accompanying the spread of matter over 
an equipotential of a rapidly rotating star take place. However, this 
all affects the instantaneous equilibrium of a neutron star only through 
global changes in its baryonic mass and angular momentum. It is 
pertinent to recall here that, according to Huygens's theorem (see, 
e.g., Tassoul 1978), the work done by gravitational and centrifugal 
forces on a particle to move it over the surface of a rigidly rotating 
liquid body in its own gravitational field is zero.

{\it Corollary 2.\quad Since the kinetic energy $\dot{M} v^2/2$ of the decelerated 
matter is released during spherical accretion onto a nonrotating cool 
star at rate $\dot{M}$ in its surface layer in unit time, there is no volume 
energy release in this case either.}

This assertion follows from theorem 2, because, in this case, $l = \Omega= 0,\, \Phi =\Phi_e =\,GM/R$, and the total energy release given by (13) 
matches the surface one.

{\bf Disk Energy Release.} Passing from one Keplerian orbit to another with 
a smaller radius, an accreting particle loses its energy through viscous 
friction, which radiates away. The total energy radiated away by one 
particle from an infinitely distant point before it approaches a 
Keplerian circular orbit on the stellar surface is equal to the particle 
energy in this orbit, with an opposite sign. Therefore, the following 
formula holds for the energy release of the entire disk:
\begin{equation}
L_d =- \dot{M}(R^2 \Omega_K^2/2 -\Phi_e).
\end{equation}
Setting the sum of centrifugal and gravitational forces equal to zero 
yields
\begin{equation}
R \Omega_K^2 = -\f{d\Phi}{dr}|_{r=R}.
\end{equation}
The formula for the disk energy release follows from (15) and (16):
\begin{equation}
L_d= \f{1}{2R}\dot{M}\f{d( r^2 \Phi)}{ dr}|_{r=R}.
\end{equation}
If there is no rotation, then $L_d=L_s=\dot{M} GM/2R$ (Shakura and 
Sunyaev 1973).

In order to estimate the effect of NS rotation on the disk energy 
release, let us consider an example in which the stellar matter is 
modeled by an ideal, incompressible fluid. In this case, the 
axisymmetric equilibrium figure is a Maclaurin spheroid. The Dirichlet formula (see, e.g., Lamb 1947) can be used for the gravitational 
potential of a homogeneous ellipsoid, and the following expression can 
be derived for $L_d$ from (17) (below, $e$ denotes the spheroid 
eccentricity, $e \equiv\sqrt{ 1-c^2/a^2},\quad c,a$ to be the smaller and
larger semiaxes):
\begin{equation} 
L_d=\dot{M}\f{3 G M}{2a}\left(\f{e^2-1}{e^3}\arcsin{e}+\f{\sqrt{1-e^e}}{e^2}\right).
\end{equation}
The Keplerian equatorial rotation frequency can be calculated from (16):
\begin{equation} 
 f_K=\sqrt{ G\rho/2\pi}B(e);\quad     B(e)\equiv\sqrt{\f{
\sqrt{1-e^2}}{e^3}\arcsin   e   -\f{1-e^2}{e^2}}.   
\end{equation}
The formula for the rotation frequency is (Lamb 1947)
\begin{equation}  f = \sqrt{ G\rho  g(e)/ 2\pi};\quad  g(e)\equiv  \f{
\sqrt{1-e^2}}{e^3}(3-2 e^2)\arcsin e -3\f{1-e^2}{e^2}. \end{equation}
According to corollary 1 of theorem 2 and formulas (19) and (20), the 
energy release on the NS surface is
\begin{equation}
L_s = \frac12 \dot{M} (\Omega_K - \Omega)^2 R^2=\dot{M}\f{3 G M}{4
R}\left(\sqrt{\f{\arcsin{e}}{e^3}-
\f{\sqrt{1-e^2}}{e^2}}-\sqrt{\f{3-2e^2}{e^3}\arcsin{e}-3\f{\sqrt{1-e^2}}{e^2}}\right)^2.
\end{equation}
For the case of counterrotation one should change the sign before the
second therm in round squares of expression (21).

Using (18 - 21), we can obtain the following unexpectedly simple 
relation to estimate the fraction of radiation from the disk in the NS 
total radiation:
\begin{equation}
L_d/(L_s + L_d) =0.5( 1+ f/ f_K).
\end{equation}
This formula is valid both for the positive values of $f$ (the case of
corotation) and for negative ones (the case of counterrotation).

Note that the eccentricity $e\rightarrow 1$ as $ f\rightarrow f_K$ for an incompressible 
fluid. Therefore, the spheroid asymptotically transforms into a plane 
disk. However, large angular velocities cannot be reached: Maclaurin 
spheroids become unstable to quadrupole perturbations at $e > 0.8127$ (at 
 $f >0.4326 \sqrt{G\rho/2\pi}$) (Lamb 1947; Chandrasekhar 1973).

According to (18) and (21), the total luminosity $ L_s + L_d $ in the 
stability range of Maclaurin spheroids satisfies the inequalities
$$
 0.0485<\f{ L_s + L_d}{ \dot{M}c^2} \left(\f{M}{1.4
M_{\odot}}\right)^{-2/3}\left(\f{\rho}{10^{14}\mbox{g/cm}^3}\right)^{-1/3}<0.11
$$
in the case of corotation and the inequalities
$$
 0.11<\f{ L_s + L_d}{ \dot{M}c^2} \left(\f{M}{1.4
M_{\odot}}\right)^{-2/3}\left(\f{\rho}{10^{14}\mbox{g/cm}^3}\right)^{-1/3}<0.1953
$$
in the case of counterrotation. The speed of light (constant) was 
introduced here for convenience of comparing the results in the 
Newtonian approximation and in general relativity (Sibgatullin and 
Sunyaev 2000).

For the critical $e = 0.8127,$ the ratio of angular velocities $\Omega/\Omega_K$ is 
0.602. It thus follows from formula (22) (in the stability range of 
Maclaurin spheroids) for spheroid and disk corotation that $0.199 < 
L_s/(L_s + L_d) < 0.5.$ For a nonrotating star 
as was shown by Shakura and Sunyaev 1973 the ratio $L_s/(L_s + L_d)$ is
equal 0.5. The inequality $0.5 < L_s/(L_s + L_d) < 0.801$ 
holds for disk and spheroid counterrotation.

\section{Treatment in General Relativity.}
The metric of stationary, axisymmetric spaces invariant to the change 
$t,\phi \Rightarrow -t,\,-\phi$  can be written in cylindrical $r,\,z,\,\phi$ coordinates as
\begin{equation}
ds^2 = a^2 \,dt^2 - b^2\, (d\phi - \omega dt)^2 -
e^{2\sigma} \,(dr^2 + dz^2),
\end{equation}
where $a,\,b,\,\sigma\,$ are the sought-for functions of $r$ and $z.$ 
The gravitational-mass functional in the steady-state, axisymmetric case 
is (Hartle and Sharp 1967)
\begin{equation}
M c^2 =c^4 \int_{\infty}\frac{R}{16\,\pi G}\sqrt{-g}\,d\phi dr\,dz +
c^{-1} 
\int_{V} T^0_0\sqrt{-g}d\phi,\,dr\, dz,
\end{equation}
where $V$ and $\partial V$ are the region occupied by the star and its boundary, respectively. The integral of the scalar curvature is extended to the 
entire three-dimensional space:
$$
T^0_0 = (p+\epsilon)\frac{(a^2+\omega\Omega b^2)}{a^2
-(\omega-\Omega)^2 b^2} -\, p;\quad \sqrt{-g} = a\,b\,e^{2\sigma}.
$$
The NS nuclear overcompressed matter is an ideal isentropic gas with a 
given internal energy density, $\epsilon =\rho u(\rho)$ , with $ d\epsilon = (p+\epsilon) d\rho/\rho.$
According to formula (24), the mass functional for a steady-state, 
axisymmetric star is determined by the functions $\rho,\, a,\,b,\,$ and $\sigma$,  of independent variables $r$ and $z$ and by the constant 
$\Omega.$ Let us consider two close equilibrium states $A$ and $A + \delta A.$ 
Subtracting the mass functional in state $A$ from the mass functional in 
state $A + \delta A$ yields
$$
\delta M =2\pi \int\left(\delta\rho\frac{\delta M}{
\delta\rho}+\delta\sigma\frac{\delta M}{   
\delta\sigma}+\delta a\frac{\delta M}{   
\delta a}+\delta b\frac{\delta M}{   
y
\delta b}\right) dr\,dz + \delta \Omega\frac{\partial M}{\partial\Omega}
+c^{-3}\int_{\partial V}T_0^0 W\,\sqrt{-g}\,dl.
$$
The functions of the form  $\frac{\delta M}{\delta a}$ are variational derivatives of the 
functional $M$ with respect to the corresponding function $a(r, z).$

Since the metric coefficients and their first derivatives are continuous 
at the boundary of the region $\partial V $, the surface terms from variations of 
the gravitational mass reduce to the integral over a distant surface 
at pseudo-Euclidean infinity, where they vanish. The variational 
derivatives of $M$ with respect to the metric coefficients are zero by 
virtue of the Einstein equations.

Using the equalities $u^0 u_0+u^{\phi} u_{\phi}= 1,$ and $ u^{\phi}=
\Omega u^0$, one could derive the 
following Hartle-Sharp formula at fixed metric coefficients:
\begin{equation}
\delta u^0 = -u_{\phi}(u^0)^2\delta \Omega.
\end{equation}
If we vary the gravitational-mass functional (24) at given metric 
coefficients, then we obtain
\begin{equation}
\delta M c^2 = 2\pi c^{-1}\delta\int_{V}T_0^0\sqrt{-g}\,dr\,dz = 2\pi c^{-1}\delta\int_{V}(\epsilon-T^0_{\phi}\Omega)\sqrt{-g}dr\,dz.
\end{equation}
We now use the expressions for rest mass $m$ and angular momentum $J$ 
(Hartle and Sharp 1967)
$$
m =2\pi\int_{V}\rho u^0\sqrt{-g}\,dr\,dz,
$$
$$
J = -2c^{-1}\pi\int_{V}T^0_{\phi}\sqrt{-g}dr\, dz
$$
and the relativistic integral of dynamical equilibrium conditions
$$
\frac{p+\epsilon}{(\rho c  u^0)} =\mu  =\mbox{const}.
$$
In view of (24), the following formula holds:
$$\delta\epsilon  = \f{p+\epsilon}{\rho}\delta \rho =\f{p+\epsilon}{\rho u^0} (\delta (\rho
u^0) -\rho \delta u^0) = \f{p+\epsilon}{\rho u^0}\delta(\rho u^0) + T^0_{\phi}\delta\Omega. 
$$
The quantity $\delta M$ can then be easily represented after reducing similar 
terms in (26) as
\begin{equation}
\delta M\,c^2 = \Omega \,\delta J + \mu\delta\, m\,
\end{equation} 
hence
\begin{equation}
\Omega=\f{\p
M c^2}{\p J}|_{m},\quad \mu = \f{\p
M\,c^2}{\p m}|_{J}
\end{equation}
{\it Theorem 3. \quad For any two close isentropic equilibrium states of a NS, 
the variations in gravitational mass, angular momentum, and rest mass 
are related by (28).}

{\bf Energy Release during Disk Accretion onto a NS in General Relativity.} 
Let us now use theorem 3 to calculate the NS radiation and evolution 
from equilibrium states. The constant in the relativistic case is also 
convenient to estimate from its value at the stellar equator:
\begin{equation}
\mu = c/u^0 = c\sqrt{a^2-(\omega-\Omega)^2 b^2}
,
\end{equation}
the metric coefficients (23) are taken at the stellar equator.

In general relativity, the NS equatorial radius can be larger and 
smaller than the radius of the marginally stable orbit, which is equal 
to three gravitational radii in the absence of rotation. In the former 
case, we denote the energy and angular momentum of a test particle of 
unit mass in a Keplerian orbit with the NS radius by $e$ and $l,$ 
respectively, take into account the law of conservation of angular 
momentum $dJ = l d m$, and equate the expressions for the change in 
gravitational mass from theorem 3 and from the law of conservation of 
energy:
$$
(\Omega l+\mu )\, dm = dm\, e - L_s dt.
$$
We use the equalities $e =c \, (u_K)_0,\quad l=c\,(u_K)_{\phi},\quad \mu =
c/u^0,$ and $ u^{\phi} = \Omega u^0$,  and express $L_s$ as
\begin{equation}
L_s=\dot{ M}(e -\Omega l -\mu) = \mu({\bf u}\cdot{\bf u}_K-1)\dot{ M},\qquad \frac{d m}{dt} = \dot{ M}.
\end{equation}
Here, we introduced the designation $ \, \cdot \, $ for a scalar product of the
4-velocity vector of a particle rotating with the NS angular velocity 
and the 4-velocity vector of a particle rotating in a Keplerian orbit at 
the same point on the stellar equator. Expression (30) for $L_s$ is always 
positive and becomes zero if the NS angular velocity reaches the 
Keplerian one at the stellar equator. In our previous paper (Sibgatullin 
and Sunyaev 2000), we gave approximation formulas for the dependences of 
$M,\, e,\, l,$ and $\, f_K$, on Kerr parameter and rest mass for particles in 
equatorial Keplerian orbits for a NS with EOS A and FPS. We used (28) to 
calculate the NS angular velocity and chemical potential. For stars with 
EOS A and FPS and with a gravitational mass of 1.4 $M_{\odot}$, we derived 
approximation formulas for the dependences of total luminosity $L_s + L_d$  
and $ L_s/(L_s+L_d)$ ratio on NS angular velocity by using (30).

The case where the NS radius is smaller than the radius of the 
marginally stable orbit is characteristic of general relativity alone. 
In this case, the gravitational field does an additional work on the 
spiraling-in particles in the gap; no stable Keplerian orbits exist 
within the marginally stable orbit. Thus, the rate of energy release is
\begin{equation}
L_s=\dot{ M}(e_* -\Omega l_* -\mu)=\mu({\bf u}\cdot{\bf u}_*-1)\dot{ M},
\end{equation}
where $ {\bf u}_*$  denotes the 4-velocity vector of the particle that fell on the 
stellar surface from a circular marginally stable orbit with the energy 
$e_*$ and angular momentum $l_*$ corresponding to this orbit. In 
paper Sibgatullin and Sunyaev (2000), we gave universal (valid for an 
arbitrary equation of state) approximation formulas for the dependences 
of $ e_*,\, l_*,\, R_*,$ and $\, f_K^*$ on Kerr parameter and on dimensionless 
quadrupole coefficient $b$ for the particles falling on the surface from 
circular marginally stable orbits. We constructed plots of total 
luminosity against angular velocity for a NS with gravitational mass $M 
= 1.4 M_{\odot}$ by using (27) and (30).

Note the fundamental difference between our formulas (30), (31) for the 
luminosity and formulas (16), (18) from Thampan and Datta (1998). These 
authors calculated the energy release produced by particles during 
accretion as a difference between the particle energy in the stable 
circular equatorial orbit nearest the body and the energy of the 
particles lying on the equator and corotating with the body.

{\bf Using Theorem 3 to Calculate the NS Angular Velocity and Equatorial 
Radius.} As the NS parameters, we choose its rest mass and Kerr 
parameter $j$ (dimensionless angular momentum $cJ/GM^2$). The methods of 
constructing the function $M(j, m)$ were developed previously 
(Sibgatullin and Sunyaev 2000).

According to (28), the NS angular velocity at given rest mass $m$ and 
Kerr parameter $j$ can be determined from the known function $M(j, m)$ by 
using the formula
$$
\Omega GM/c^3 = \f{M,_{j}|_m}{M+2j M,_{j}|_m}.
$$ 
Another important formula that relates the constant $\mu$ to a 
derivative of the gravitational mass with respect to the rest mass at 
constant angular momentum follows from (28). Taking (29) for $\mu$, we 
obtain
\begin{equation}
c\f{M M,_{m}|_j}{M + 2j M,_{j}|_m} =\sqrt{ a^2 - b^2\,(\Omega-\omega)^2},
\qquad R  = b,
\end{equation}
where $R$ is the geometric equatorial radius of the star (the equator 
length divided by 2@[pi]); the metric coefficients $a$ and $b$ are taken 
at the NS equator.

In the static case, the NS radius can be determined from (32) by using 
the Schwarzschild metric ($\omega = \Omega = 0, a^2 =c^2\,( 1- \f{2 G M}{R
c^2}$):
\begin{equation}
\f{R c^2}{GM}=\f{2}{1-M,_{m}^2}.
\end{equation}
Remarkably, formula (33) [if $M(j, m)$ is substituted in it and 
differentiated at constant $j$] closely agrees with the numerical data 
from Cook et al. (1994) and Stergioulus (1998) for the equatorial 
radius of a rotating NS, to within a few percent:$
R\approx 2G M/(c^2-c^2 M,_{m}^2|_j).
$

ACKNOWLEDGMENTS

We wish to thank N.Stergioulus for the use of his numerical code, who made 
it available in the Internet. We express our gratitude to Yu. Astakhov for
the help of translating  this paper in English.

REFERENCES

1. M. A. Alpar, astro-ph/991228.

2. J. M. Bardeen, Astrophys. J. v.162, 71 (1970).

3. G. Biehle and R. D. Blanford, Astrophys. J. v.411, 302 (1993).

4. S. Chandrasekhar, Ellipsoidal Figures of Equilibrium (Yale Univ. 
Press, New Haven, 1969; Mir, Moscow, 1973).

5. G. G. Chernyi, Gas Dynamics (Nauka, Moscow, 1988).

6. G. B. Cook, S. L. Shapiro, and S. A. Teukolsky, Astrophys. J. v.424, 
823 (1994).

7. L. Crocco, Zeitschrift fuer Angewandte Math. und Mech. (ZAMM) v.17, 1 (1937).

8. J. B. Hartle and D. H. Sharp, Astrophys. J. v.147, 317 (1967).

9. N. A. Inogamov and R. A. Sunyaev,  
Astronomy Lett. v.25, 269 (1999); astro-ph/9904333.

10. W. Kley, Astron. Astrophys. v.247, 95 (1991).

11. W. Kluzniak, Ph.D. Thesis (Stanford Univ., 1987).

12. W. Kluzniak and R. V. Wagoner, Astrophys. J. v.297, 548 (1985).

13. 7. H. Lamb, Hydrodynamics (Cambridge Univ. Press, Cambridge, 1932; 
Gostekhizdat, Moscow, 1947).

14. L. D. Landau and E. M. Lifshitz, Statistical Physics (Nauka, Moscow,
1976; Pergamon, Oxford, 1980), Part 1.

15. K. P. Levenfish, Yu. A. Shibanov, and D. G. Yakovlev, Pis'ma Astron. 
Zh. v.25, 417 (1999) [Astron. Lett. 25, 417 (1999)].

16. M. C. Miller and F. K. Lamb, Astrophys. J. v.470, 1033 (1996).

17. J. P. Ostriker and J. E. Gunn, Astrophys. J. v.157, 1395 (1969).

18. K. Oswatitsch, The Foundations of Gas Dynamics, (Springer Verlag, 
New York, 1976).

19. R. Popham and R. Narayan, Astrophys. J. v.442, 337 (1995).

20. N. I. Shakura and R. A. Sunyaev, Astron. Astrophys. v.24, 337 (1973).

21. N. P. Sibgatullin and R. A. Sunyaev, Astronomy Lett. v.24, 774 (1998)].

22. N. P. Sibgatullin and R. A. Sunyaev, Astronomy Lett. v. 26, 699 ( 2000) (astro-ph/0011253).

23. N. Stergioulus, livingreview.org/Articles/Volume 1/1198-8stergio.

24. J. L. Tassoul, Theory of Rotating Stars (Princeton Univ. Press, 
Princeton, 1978).


 \end{document}